\documentclass[sigconf, nonacm]{acmart}
\usepackage{bm}
\usepackage{enumitem}
\usepackage{booktabs}
\usepackage{adjustbox}
\usepackage{multirow}
\usepackage{svg}
\usepackage{graphicx} 
\usepackage{makecell}

\newcommand\vldbdoi{XX.XX/XXX.XX}
\newcommand\vldbpages{XXX-XXX}
\newcommand\vldbvolume{14}
\newcommand\vldbissue{1}
\newcommand\vldbyear{2020}
\newcommand\vldbauthors{\authors}
\newcommand\vldbtitle{\shorttitle} 
\newcommand\vldbavailabilityurl{URL_TO_YOUR_ARTIFACTS}
\newcommand\vldbpagestyle{plain} 

\newcommand{\vpara}[1]{\vspace{0.05in}\noindent \textbf{#1 }}
\NewDocumentCommand{\haoyang}{mO{}}{\textcolor{blue}
{\textsuperscript{\textit{haoyang}}\textsf{\textbf{\small[#1]}}}}

\begin{document}

\title{Is Large Language Model Good at Database Knob Tuning? A Comprehensive Experimental Evaluation}

\author{
    Yiyan Li$^{1\ast}$,
    Haoyang Li$^{1\ast}$,
    Zhao Pu$^1$,
    Jing Zhang$^1$,
    Xinyi Zhang$^1$,
    Tao Ji$^1$,
    Luming Sun$^2$,
    Cuiping Li$^1$,
    Hong Chen$^1$
}
\affiliation{
  $^1$ School of Information, Renmin University of China, $^2$ Shanghai Yunxi Technology Co., Ltd
  \country{China}
}

\email{
  {liyiyan, lihaoyang.cs, puzhao, zhang-jing, xinyizhang.info, jitao, licuiping, chong}@ruc.edu.cn
}
\email{
  sunluming@inspur.com 
}

\begin{abstract}
Knob tuning plays a crucial role in optimizing databases by adjusting knobs to enhance database performance. However, traditional tuning methods often follow a Try-Collect-Adjust approach, proving inefficient and database-specific. Moreover, these methods are often opaque, making it challenging for DBAs to grasp the underlying decision-making process. 

The emergence of large language models (LLMs) like GPT-4 and Claude-3 has excelled in complex natural language tasks, yet their potential in database knob tuning remains largely unexplored. This study harnesses LLMs as experienced DBAs for knob-tuning tasks with carefully designed prompts. We identify three key subtasks in the tuning system: knob pruning, model initialization, and knob recommendation, proposing LLM-driven solutions to replace conventional methods for each subtask. 

We conduct extensive experiments to compare LLM-driven approaches against traditional methods across the subtasks to evaluate LLMs' efficacy in the knob tuning domain. Furthermore, we explore the adaptability of LLM-based solutions in diverse evaluation settings, encompassing new benchmarks, database engines, and hardware environments. Our findings reveal that LLMs not only match or surpass traditional methods but also exhibit notable interpretability by generating responses in a coherent ``chain-of-thought'' manner. We further observe that LLMs exhibit remarkable generalizability through simple adjustments in prompts, eliminating the necessity for additional training or extensive code modifications.

Drawing insights from our experimental findings, we identify several opportunities for future research aimed at advancing the utilization of LLMs in the realm of database management.
\end{abstract}


\maketitle

\pagestyle{\vldbpagestyle}
\begingroup\small\noindent\raggedright\textbf{PVLDB Reference Format:}\\
\vldbauthors. \vldbtitle. PVLDB, \vldbvolume(\vldbissue): \vldbpages, \vldbyear.\\
\href{https://doi.org/\vldbdoi}{doi:\vldbdoi}
\endgroup
\begingroup

\renewcommand\thefootnote{}\footnote{\noindent
$^\ast$Yiyan Li and Haoyang Li contribute equally to this paper. \\
\noindent This work is licensed under the Creative Commons BY-NC-ND 4.0 International License. Visit \url{https://creativecommons.org/licenses/by-nc-nd/4.0/} to view a copy of this license. For any use beyond those covered by this license, obtain permission by emailing \href{mailto:info@vldb.org}{info@vldb.org}. Copyright is held by the owner/author(s). Publication rights licensed to the VLDB Endowment. \\
\raggedright Proceedings of the VLDB Endowment, Vol. \vldbvolume, No. \vldbissue\ %
ISSN 2150-8097. \\
\href{https://doi.org/\vldbdoi}{doi:\vldbdoi} \\
}\addtocounter{footnote}{-1}\endgroup

\ifdefempty{\vldbavailabilityurl}{}{
\vspace{.3cm}
\begingroup\small\noindent\raggedright\textbf{PVLDB Artifact Availability:}\\
The source code, data, and/or other artifacts have been made available at \url{https://github.com/intlyy/Knob-Tuning-with-LLM}.
\endgroup
}

\section{Introduction}

Configuration knobs control many aspects of database systems (\emph{e.g.}, memory allocation, thread scheduling, caching mechanisms), and different combinations of knob values significantly affect performance, resource usage, and robustness of the database~\cite{Chaudhuri2007@selftuningsystem}. In general, given a workload, knob tuning aims to judiciously adjust the values of knobs to improve the database performance~\cite{Zhao23@tuningsurvey}. For example, the MySQL database has about 260 knobs, of which adjusting the \textit{InnoDB\_buffer\_ pool\_size} and \textit{tmp\_table\_size} can significantly improve database query processing efficiency~\cite{Li19@qtune}. Therefore, it is vital to set proper values for the knobs.

\begin{figure}
    \centering
    \includegraphics[width=0.47\textwidth]{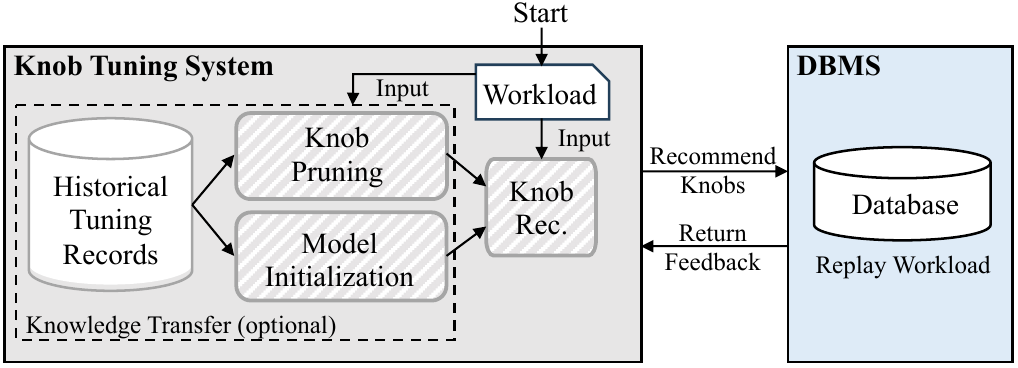}
    \caption{Overview of a knob tuning system. ``Knob Rec.'' stands for ``Knob Recommendation''. Knob pruning and model initialization serve as optional components within the system, designed to expedite the tuning process of the knob recommendation methods.} 
    \label{fig:overview}
\end{figure}

Traditional knob tuning relies on database administrators (DBAs) to manually try out typical knob combinations based on their experience. This process is labor-intensive and impractical for a large number of database instances (\emph{e.g.}, tens of thousands on cloud platforms)~\cite{Pavlo21@self-drivingDB}. Leveraging machine learning techniques, researchers have developed various automated knob tuning systems capable of identifying suitable knob values without human intervention~\cite{Aken17@OtterTune, Li19@qtune, Kanellis22@llamatune, Lin22@SparkTune, Tan19@iBTune}. The workflow of these systems is depicted in Figure~\ref{fig:overview}, with the left side illustrating the tuning system and the right side representing the target database management system (DBMS). Upon receiving a workload, the tuning method suggests a configuration for the DBMS, which is then tested with the workload to measure performance metrics (\emph{e.g.}, latency or transactions per second). Subsequently, based on this feedback, the tuning method refines its policy and proposes a new configuration. Through multiple iterations of ``Try-Collect-Adjust'', an optimized configuration can be achieved to enhance database performance significantly under the given workload.

The knob tuning system can be segmented into three key components: \textit{knob recommendation}, \textit{knob pruning}, and \textit{model initialization}. \textit{Knob recommendation} serves as the cornerstone of the tuning system, offering suggestions for suitable configurations tailored to the workload. The approaches for knob recommendation fall into four primary categories: reinforcement learning (RL) based approaches~\cite{Cai22@Hunter, Li19@qtune, Zhang21@CDBTune, Trummer22@DBBert}, Bayesian optimization (BO) based techniques~\cite{Duan09@iTuned, Zhang21@ResTune, Aken17@OtterTune, Zhang22@OnlineTune}, deep learning (DL) based methods~\cite{Tan19@iBTune, Aken21@deeplearning-tuning, Lin22@Sparktuning}, and heuristic methods~\cite{Zhu17@BestConfig, Chen11@heuristic-method1}. Given the expansive search space of configurations, these tuning methods typically necessitate numerous interactions with the DBMS, with each iteration involving workload execution. This iterative process is both time and resource-intensive. To address this challenge, various knowledge transfer methods~\cite{Cereda21@CGPTune, Aken17@OtterTune, Li19@qtune, Zhang21@ResTune, Saltelli02@SA} have been introduced, leveraging past tuning records to expedite the tuning process. These methods can be classified into two categories: \textit{knob pruning} and \textit{model initialization}. \textit{Knob pruning} targets the selection of crucial knobs and the determination of their reasonable ranges for the specific workload, thereby reducing the configuration space~\cite{Duan09@iTuned, Saltelli02@SA, David04@Softwaretune, Tan19@iBTune, Tibshirani96@lasso, Kanellis20@TooManyKnobs, Debnath08@SARD}. On the other hand, \textit{model initialization} focuses on initializing the learnable model within the knob recommendation methods, which can accelerate their convergence speed~\cite{Zhang21@ResTune, Aken17@OtterTune, Li19@qtune, Feurer2018@RGPE, Zhang21@CDBTune, Cereda21@CGPTune, Ge21@WATuning}. It should be noted that the knob pruning and model initialization techniques usually occur at the beginning of the tuning phase, which are optional components of the tuning system. Then, the knob recommendation methods iteratively interact with DBMS until the database performance coverage or stop conditions are triggered.

\vpara{Limitations of Existing Methods.} Despite the notable performance achieved by current methods, they still exhibit the following limitations. (1) Knob pruning and model initialization techniques often heavily rely on historical tuning data or domain knowledge (\emph{e.g.}, database manual, and forum discussions) to expedite current tuning tasks. For instance, \textit{knob pruning} methods like Lasso~\cite{Tibshirani96@lasso} and Sensitivity Analysis~\cite{Nembrini18@SA} necessitate extensive historical tuning data for calculating knob importance rankings and GPTuner~\cite{lao2024@gptuner} and DB-BERT~\cite{Trummer22@DBBert} requires manually collected knob-tuning-related texts to optimize the configuration space. Similarly, \textit{model initialization} methods like QTune~\cite{Li19@qtune} also rely on historical data for pre-training actor and critic models. However, acquiring such data can be costly, particularly when addressing new database kernels or hardware environments, requiring data collection from scratch. (2) Regarding knob recommendation methods, many of them need to replay the workload in each iteration to capture performance metrics. Due to the limited exploration and exploitation capabilities of these methods, they often require numerous iterations, leading to significant time and resource expenses. (3) Almost all database knob tuning approaches operate as black boxes. This opacity makes it challenging for DBAs to understand the rationale behind recommended outcomes and complicates their ability to intervene effectively in case of issues.

\vpara{Our Proposal.} This paper aims to explore the feasibility of utilizing LLMs to emulate the behaviors of DBAs in performing knob-tuning-related subtasks, including knob pruning, model initialization, and knob recommendation. Recent advancements in LLMs have yielded remarkable breakthroughs in diverse domains, such as mathematical reasoning~\cite{ahn2024@math}, text-to-SQL~\cite{li2024@codes}, and tool using~\cite{qin2023@toolllm}. LLMs are famous for vast knowledge, strong reasoning capabilities, and remarkable interpretability, offering potential solutions to the aforementioned limitations within the tuning system. Therefore, integrating LLMs into the database knob tuning system represents a promising direction for research. While LLM-based tuning methods like GPTuner~\cite{lao2024@gptuner} have been proposed, existing work primarily focuses on knob pruning, only one subtask within the broader knob tuning process.




In this study, we carefully craft prompts for each tuning subtask and evaluate LLMs' performance through comparative experiments against previous state-of-the-art (SOTA) methods. Given the diverse array of LLMs available, our evaluation extends beyond a single model. We explore a spectrum of powerful LLMs, including GPT-3.5~\cite{ouyang2022@instructgpt}, GPT-4-Turbo~\cite{openAI23@GPT4}, GPT-4o~\cite{openai2024@gpt4o}, and Claude-3-Opus~\cite{Claude_3}. However, as these powerful LLMs are closed-sourced, concerns related to data privacy and high usage costs may arise. To address this, we additionally consider several open-source LLMs, such as LLaMA3~\cite{meta2024@llama3} and Qwen2~\cite{alibaba2024@qwen2}, which offer the advantage of local deployment.

Our primary experiments are conducted using an Online Transaction Processing (OLTP) benchmark (SYSBENCH~\cite{sysbench}) in conjunction with the MySQL database engine. In addition, given the inherent adaptability of LLMs, which allows them to generalize to new scenarios through prompt adjustments, we also conduct comprehensive assessments to evaluate the generalizability of our LLM-based solutions across diverse workloads, database engines, and hardware environments. We believe that this study can serve as a source of inspiration for more AI4DB tasks, such as query optimization and index recommendation.

We make the following contributions in this paper:
\begin{itemize}[leftmargin=1em]
\setlength\itemsep{0em}
\item We investigate the capabilities of LLMs in executing three knob tuning subtasks: knob pruning, model initialization, and knob recommendation. For each subtask, we carefully craft prompts to guide the LLMs in effectively addressing the specific objectives. 

\item In our experiments, we comprehensively evaluate both closed-source and open-source LLMs, offering researchers and practitioners a thorough understanding of the strengths and limitations of various LLMs.



\item We additionally assess the generalizability of LLMs by conducting experiments across various benchmarks, database engines, and hardware environments.

\item Based on our findings, we explore future research directions and potential challenges in the domain of utilizing LLMs for knob tuning.

\end{itemize}

The remainder of the paper is organized as follows. We formally define the problems in Section~\ref{sec:problem definition}, followed by a description of the integration of LLMs with three database knob tuning subtasks in Section~\ref{sec:tuning with LLM}. Then, Section~\ref{sec:setup}-\ref{sec: generalization} presents our experimental evaluation and main findings. Finally, we discuss research opportunities in Section~\ref{sec:discussion} and conclude in Section~\ref{sec:conclusion}.

\section{PROBLEM DEFINITION}
\label{sec:problem definition}

Consider a modern database system equipped with $m$ tunable knobs, represented as \(\theta_1, ..., \theta_m\). Each knob \(\theta_i\) might be either continuous or categorical, covering a range of configurable database aspects like work memory size and maximum connection limits. Every knob \(\theta_i\) is assigned a value within a predetermined range \(\Theta_i\), signifying the allowable value spectrum for that knob. The combination of possible knob values forms a huge multi-dimensional configuration space, represented by \( \bm{\Theta}  = \Theta_1 \times \Theta_2 \times ... \times \Theta_m\). A specific point within this space signifies a unique database configuration, characterized by a set of knob values \(\bm{\theta} =  (\theta_1^*, \theta_2^*, ..., \theta_m^*) \in \bm{\Theta}\).

In the context of optimizing database performance, we define the performance metric as \(f\), representing factors like throughput or latency that we seek to enhance. For a given database instance \(D\), workload \(W\), and a specific configuration \(\bm{\theta}\), the resulting performance metric \(f(D, W, \bm{\theta})\) is observed after applying \(\bm{\theta}\) in the database engine and executing \(W\) on \(D\). 

As illustrated in Figure~\ref{fig:overview}, a complete knob tuning system encompasses three important subtasks: knob pruning, model initialization, and knob recommendation. The objective of this study is to explore the ability of LLMs to execute these subtasks, prompting us to define the problem for each subtask via LLMs as follows.



\vpara{LLMs for Knob Pruning.}
In modern database systems, although there are hundreds of adjustable knobs, not all knobs are equally important under specific workloads. For example, working memory size is vital to memory-intensive workloads, maximum IO concurrency is vital to IO-intensive workloads. Hence, considering the characteristics of $D$ and $W$, the goal of knob pruning is to identify the most impactful knobs and define their crucial ranges. By reducing the search space, knob tuners can concentrate on adjusting these selected knobs in the constrained ranges and thus streamline the tuning process. Formally, we have:
\begin{equation}
\begin{split}
    LLM(Prompt_{kp}, D, W, \{\theta_1, ..., \theta_m\}, & \{\Theta_1, ..., \Theta_m\}) \rightarrow \\
    & \{\theta_j, ..., \theta_k\}, \{\Theta_j^{'}, ..., \Theta_k^{'}\},
\end{split}
\end{equation}
where $LLM(\cdot)$ denotes the large language model, $Prompt_{kp}$ represents the pre-defined prompt used for the knob pruning task, and the outputs $\{\theta_j, ..., \theta_k\}$ and $\{\Theta_j^{'}, ..., \Theta_k^{'}\}$ represent the LLM-selected significant knobs and their respective important ranges. 
Notably, unlike traditional knob pruning methods such as Lasso~\cite{Tibshirani96@lasso} and Sensitivity Analysis~\cite{Nembrini18@SA}, which only select knobs, the LLM-based approach can also recommend important value ranges for the selected knobs.
Furthermore, unlike Lasso~\cite{Tibshirani96@lasso} and Sensitivity Analysis~\cite{Nembrini18@SA}, which rely on historical tuning data, and the existing LLM-based method GPTuner~\cite{lao2024@gptuner} and DB-BERT~\cite{Trummer22@DBBert}, which requires manually collected knob-tuning-related texts for input augmentation, our LLM-based approach harnesses the inherent capacity of LLMs to emulate the actions of DBAs for knob pruning. Given that powerful LLMs have likely encountered tuning-related manuals and web pages during pre-training, the primary objective is to instruct them to follow knob pruning guidelines and leverage their internal knowledge. 




\vpara{LLMs for Model Initialization.}
In practical scenarios, workloads often exhibit dynamic changes, with workload pressures varying significantly over time (from morning to evening, weekdays to weekends, or workdays to holidays). 
It is widely acknowledged that tuning a specific configuration is necessary for different workloads. However, starting the tuning process from scratch for each workload requires multiple iterations of database interactions, which can be time-consuming and resource-intensive. To accelerate the tuning speed, several transfer learning-based studies have been proposed to leverage knowledge from historical tuning records as the initialization of the tuning method, facilitating quicker convergence.

Instead of accumulating extensive historical tuning data for model initialization, we propose leveraging LLMs to recommend a set of effective initial knob configurations for the new workload. These LLM-generated configurations can then be used to initialize traditional Bayesian Optimization (BO)-based tuning methods, such as HEBO~\cite{Cowen-Rivers22@HEBO} and VBO~\cite{Duan09@iTuned}. By eliminating the initial phase of random exploration, this methodology enables the BO-based methods to rapidly converge to a suitable configuration, accelerating the overall tuning process. 
Specifically, we use LLMs to sample a set of effective configurations for a given workload $W$ on database $D$:
\begin{equation}
\label{eq:model_init}
    LLM(Prompt_{rec}, \bm{\theta_{df}}, D, W, \bm{\Theta}, F_{df}) \rightarrow \{\bm{\theta}_1, ..., \bm{\theta}_u\},
\end{equation}
where \(Prompt_{rec}\) indicates the prompt used to recommend configurations, $\bm{\theta_{df}}$ represents the default configuration, \(\bm{\Theta}\) signifies the space of possible configurations, $F_{df}$ represents the database's feedback under the default configuration, and the output \(\{\bm{\theta}_1, ..., \bm{\theta}_u\}\) consists of a set of effective configurations derived from the LLM. The default configuration, denoted as $\bm{\theta_{df}}$, serves as an anchor point, guiding LLMs to adjust only the knobs requiring modification while maintaining the settings of those that do not necessitate changes. The database's feedback $F_{df}$ consists of performance metrics (such as latency or transactions per second) and internal metrics (such as lock\_deadlocks and os\_data\_writes). The feedback can provide insights into system states, enabling LLMs to identify performance bottlenecks and make necessary adjustments to the default configuration. The set of configurations $\{\bm{\theta}_1, ..., \bm{\theta}_u\}$ generated by the LLM can be used to initialize the BO-based tuning methods, serving as their starting data points.





\vpara{LLMs for Knob Recommendation.} The knob recommendation component is pivotal within the tuning system, responsible for suggesting the optimal configuration for specific workloads to enhance performance metrics. Existing techniques, including the prominent BO-based methods~\cite{Duan09@iTuned, Zhang21@ResTune, Aken17@OtterTune, Zhang22@OnlineTune} as well as RL-based approaches~\cite{Cai22@Hunter, Li19@qtune, Zhang21@CDBTune, Trummer22@DBBert}, often require hundreds of iterations to converge, hindered by their limited abilities in balancing exploration and exploitation. In this study, we posit that LLMs, with their advanced understanding of database feedback and superior exploration and exploitation capabilities, can pinpoint appropriate configurations in significantly fewer iterations. 

The LLM-based knob tuning approach is iterative: starting from the default configuration, we employ LLMs to progressively refine the configuration based on the database feedback. Formally, the refinement process is defined as:
\begin{equation}
    LLM(Prompt_{rec}, \bm{\theta_i}, D, W, \bm{\Theta}, F_{i}) \rightarrow \bm{\theta_{i+1}},
\end{equation}
here, $Prompt_{rec}$ also denotes the knob recommendation prompt, and $\bm{\theta_i}$ signifies the current configuration, $F_{i}$ represents the database feedback under current configuration, the output $\bm{\theta_{i+1}}$ represent the refined configuration. Subsequently, $\bm{\theta_{i+1}}$ is applied in the database, and the workload is executed to gather the feedback $F_{i+1}$. Then, we can start a new iteration to refine $\bm{\theta_{i+1}}$. Initially, $\bm{\theta_{0}}$ and $F_{0}$ are $\bm{\theta_{df}}$ and $F_{df}$, respectively. This refinement process iterates several times until reaching the stop criterion.

\vpara{Summary.}
In this section, we propose dividing the knob tuning tasks into three key subtasks and formulating solutions for each using LLMs. The goal is to replace traditional methods with LLM-based solutions for each subtask and evaluate their effectiveness, rather than presenting a comprehensive framework where all three subtasks are solved by LLMs. 





\section{KNOB TUNING WITH LLM}
\label{sec:tuning with LLM}
In this section, we will delve into the details of constructing prompts for three fundamental tuning subtasks. 

\subsection{Knob Pruning}
\label{sec:knob selection}
\begin{figure}
    \centering
    \includegraphics[width=0.48\textwidth]{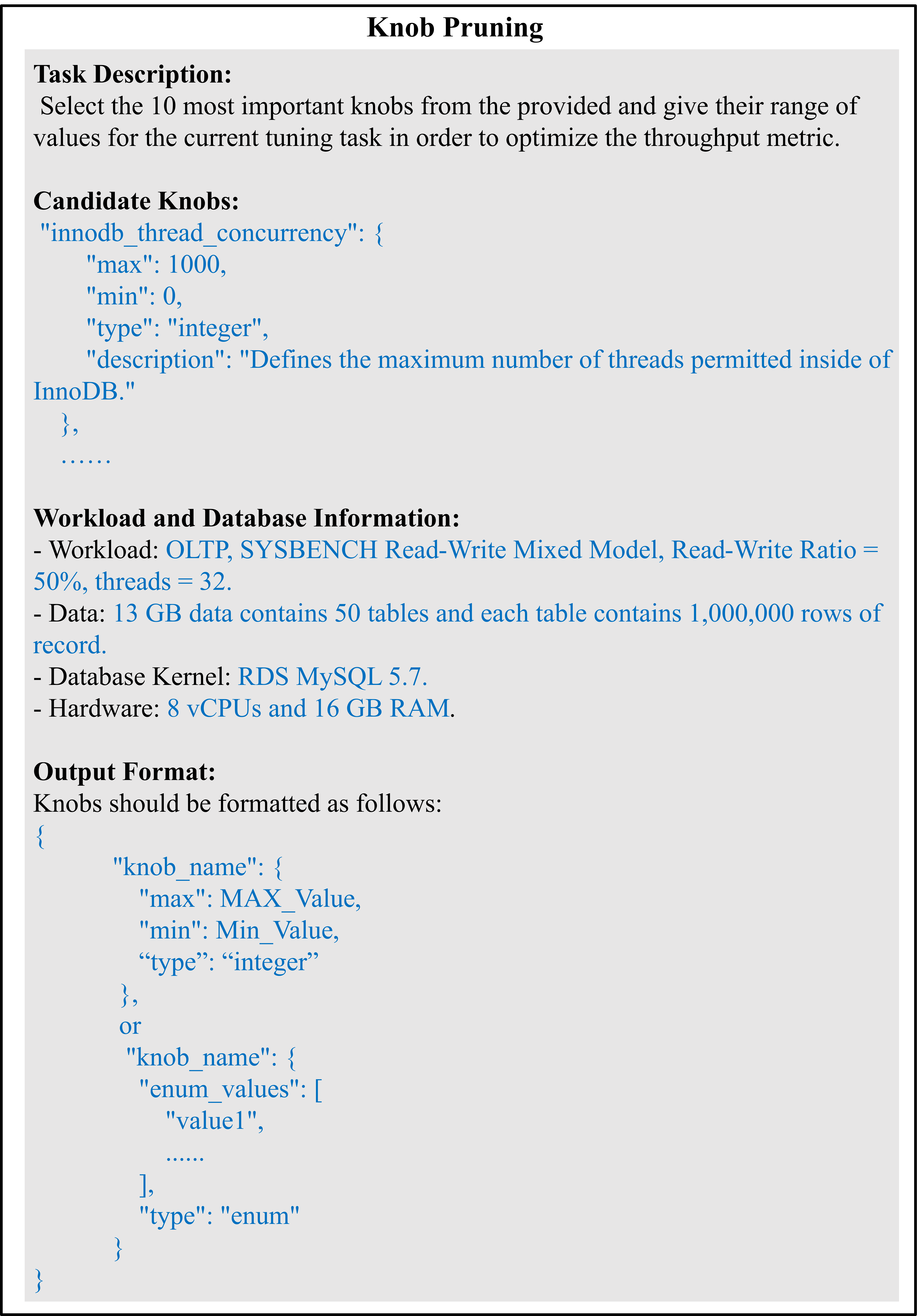}
    \caption{The prompt to perform knob selection task.} 
    \label{fig:p_selection}
\end{figure}

For a given workload, knob pruning is a critical process aimed at identifying the most important knobs and narrowing their permissive ranges, which can reduce the search space of the knob recommendation methods. Leveraging the LLM as an alternative to traditional knob pruning methods involves incorporating several important elements within the prompt. As illustrated in Figure~\ref{fig:p_selection}, the prompt for knob pruning contains the following elements:
\begin{itemize}[leftmargin=1em]
\setlength\itemsep{0em}
\item ``Task Description'' describes the objective of the LLM.
\item ``Candidate Knobs'' provides detailed information about candidate knobs within the database engine, encompassing knob names, allowable ranges, types of knobs, and their respective descriptions.
\item ``Workload and Database Information'' contains crucial details about the workload, data characteristics, database kernel, and the hardware. 
\item ``Output Format'' specifies the response format of the LLM. Specifically, it requires the LLM to enumerate the names of the chosen knobs, as well as their corresponding ranges and types, in an organized JSON format.
\end{itemize}

\subsection{Model Initialization}
\begin{figure}
    \centering
    \includegraphics[width=0.48\textwidth]{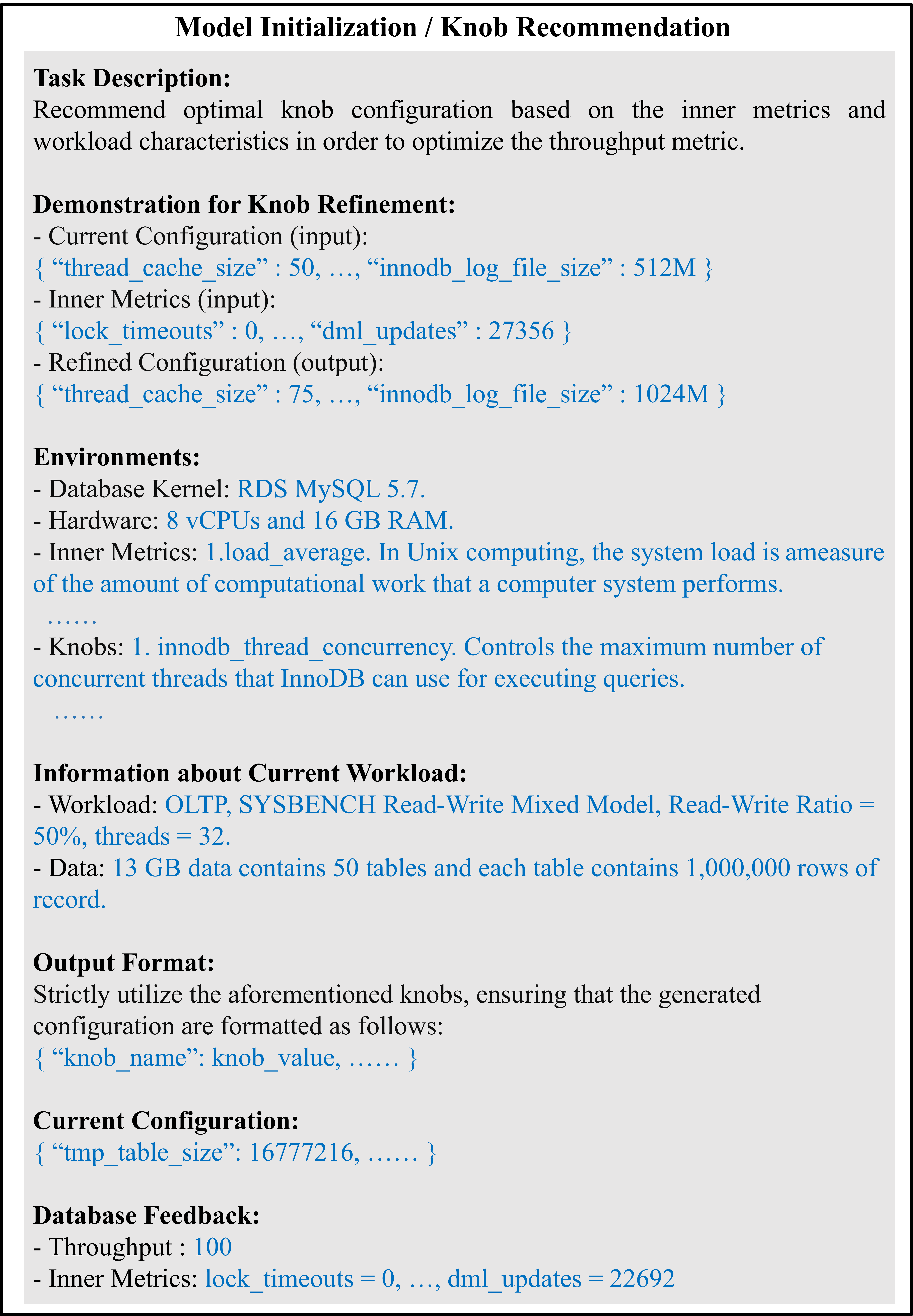}
    \caption{The prompt to perform the model initialization and knob recommendation tasks.} 
    \label{fig:p_transfer}
\end{figure}

The model initialization technique is designed to speed up the tuning process by leveraging historical tuning records to initialize the model used in the knob recommendation methods. In this paper, we concentrate on utilizing LLMs to produce a set of effective configurations for the given workload. Subsequently, these configurations can be used to initialize the BO-based methods, effectively accelerating their convergence speed. As illustrated in Figure~\ref{fig:p_transfer}, we construct the prompt for model initialization in the following format:
\begin{itemize}[leftmargin=1em]
\setlength\itemsep{0em}
\item ``Task Description'' outlines the objective of the LLM.
\item ``Demonstration for Knob Refinement'' includes a knob refinement instance, which aims to serve as the one-shot example in the prompt for demonstration purposes. This instance includes a current configuration, inner metrics, and a refined configuration.
\item ``Environment'' contains the information about database kernel and hardware information. The database kernel details encompass the database engine's name and version. The hardware information specifies the number of CPUs and the available memory resources. In addition, we also include the text descriptions of each inner metric and tunable knob.
\item ``Information about Current Workload'' includes features about the current workload, such as workload type (OLAP or OLTP) and read-write ratio, and data statistics in the database.
\item ``Output Format'' specifies the format for LLM responses.
\item ``Current Configuration'' displays the default values of the knobs, which serve as the anchor point as discussed in Section~\ref{sec:problem definition}.
\item ``Database Feedback'' showcases the performance and inner metrics of the database when executing the given workload under the default configuration. Incorporating this feedback is essential, as DBAs often depend on these metrics to assess the database's status and implement necessary adjustments. For example, confronted with a low cache hit rate, DBAs typically choose to increase the cache size to improve database performance.
\end{itemize}

In practice, we utilize the LLM for multiple samplings to acquire a collection of effective configurations. These configurations then act as the initial points for BO-based knob recommendation methods.

\subsection{Knob Recommendation}
\label{sec:recommendation}
The knob recommendation emerges as the crucial subtask within the tuning system, aimed at identifying a promising configuration for a specific workload. As elaborated in Section~\ref{sec:problem definition}, the LLM undertakes the task of knob recommendation through an iterative process. Initially starting from the default configuration, the LLM employs iterative refinements based on feedback from the database. The prompt for knob recommendation closely mirrors the prompt of model initialization, as depicted in Figure~\ref{fig:p_transfer}. The key distinction lies in the fact that, for knob recommendation, both the ``Current Configuration'' and ``Database Feedback'' are subject to change with each iteration.

\section{GENERAL SETUPS OF EVALUATION}
\label{sec:setup}
This study conducts a series of comprehensive evaluations to assess the efficacy of various LLMs across three database knob tuning subtasks. We detail the configurations of the primary experiments, encompassing hardware, software, benchmark, tuning settings, and large language models, as outlined below:



\vpara{Hardware and Software.} Our knob tuning framework is deployed across three distinct servers. The first server, dedicated to the tuning system, is equipped with 48 CPUs and 256 GB of RAM. The second server, designated for the DBMS deployment, features 8 CPUs and 16 GB of RAM, running RDS MySQL version 5.7. The third server is allocated for deploying local LLMs and is equipped with two NVIDIA A100 80GB GPUs, 80 CPUs, and 256 GB of memory. We utilize vLLM~\cite{kwon2023vLLM} as the backend to manage local LLMs. These three machines are interconnected via an intranet, communicating through a high-speed network.

The first server, functioning as the tuning system, controls coordination among three distinct servers. In our LLM-integrated knob tuning framework, the tuning system acts as a bridge between the second server (DBMS) and the third server (local LLMs), handling the interactions between them. For example, for the knob recommendation task, the tuning system first sends the default configuration to the second server to obtain feedback from the database. Then, the tuning system integrates the workload features, current configuration, database feedback, and other required information into a prompt and then sends it to the third server to obtain the LLM's response (\emph{i.e.}, refined configuration). We should note that, for closed-source LLMs, we access them through APIs, eliminating the need for using the third server.


\vpara{Benchmark.} Following previous work~\cite{Zhang22@tuning_evaluation, Zhang22@OnlineTune, Li19@qtune}, we employ SYSBENCH~\cite{sysbench}, a prevalent OLTP benchmark, for our evaluation. In particular, we focus on the OLTP-Read-Write workload within SYSBENCH, representing a workload that encompasses both read and write operations typical in OLTP scenarios. Subsequently, we load 50 tables within SYSBENCH, with each table housing 1,000,000 rows of records, culminating in approximately 13 GB of data. For a specific configuration, to conduct a stress test, we run the workload for two minutes to obtain the transactions per second (TPS) metric as the database performance. We restart the database after applying a new configuration to guarantee that all knobs have been correctly configured.

\vpara{Tuning Settings.} In the knob pruning task, both LLMs and baseline methods are tasked with identifying 10 significant knobs from the provided set of 100 candidate knobs. Following this selection, we utilize the traditional knob recommendation method, SMAC~\cite{Hutter11@SMAC}, to identify a suitable configuration using these 10 chosen knobs. In the context of model initialization and knob recommendation, we manually select 20 crucial knobs for further tuning. For model initialization, we leverage the LLM-generated configurations to initialize a BO-based knob recommendation method, VBO~\cite{Duan09@iTuned}, aiming to expedite its tuning process. For knob recommendation, we directly compare LLMs against those of traditional knob recommendation methods, including DDPG~\cite{Lillicrap15@DDPG}, SMAC~\cite{Hutter11@SMAC}, and VBO~\cite{Duan09@iTuned}.

\vpara{Evaluation Metric.} We evaluate our methods and baselines across two key dimensions: the tuning efficiency score (TES) and the optimal database performance (ODP). As illustrated in Figure~\ref{fig:overview}, we have noticed that a significant portion of the tuning process is dedicated to the DBMS side, as each iteration requires the workload to be replayed using the newly suggested configuration. To minimize the influence of external variables such as network latency, we introduce the TES metric, which quantifies the iterations needed to achieve peak database performance in the tuning process. In addition, the ODP metric measures the maximum achievable TPS during the tuning procedure.




\vpara{Large Language Models.} With the rapid advancement of LLMs, a plethora of models have surfaced, demonstrating a wide range of capabilities and applications across various domains~\cite{zhao23@LLMSurvey}. Typically, closed-source LLMs like GPT-4o and Claude-3-Opus are more powerful than available open-source models. However, closed-source LLMs come with certain limitations, including (1) concerns regarding data privacy and (2) high utilization costs. Opting for less powerful but open-source LLMs can help mitigate these issues. Therefore, to provide a comprehensive evaluation of current LLMs, we have included 4 closed-source LLMs and 3 open-source SOTA LLMs in our evaluation. Specifically, our evaluation features the following closed-source LLMs: GPT-3.5~\cite{ouyang2022@instructgpt}, GPT-4o~\cite{gpt-4o}, GPT-4-Turbo~\cite{gpt-4}, and Claude-3-Opus~\cite{Claude_3}. For publicly available LLMs, we have selected: Llama3-8B-Instruct\footnote{\url{https://huggingface.co/meta-llama/Meta-Llama-3-8B-Instruct}}~\cite{meta2024@llama3}, Llama3-70B-Instruct\footnote{\url{https://huggingface.co/meta-llama/Meta-Llama-3-70B-Instruct}}~\cite{meta2024@llama3}, and Qwen2-7B\footnote{\url{https://huggingface.co/Qwen/Qwen2-7B-Instruct}}~\cite{alibaba2024@qwen2}.

\section{Knob Pruning}
\label{sec:selection}
\subsection{Baselines}
To evaluate the knob pruning capability of LLMs, we utilize a learning-based method, SHAP~\cite{Lundberg17@SHAP}, as a competitive baseline. SHAP provides a unified framework for interpreting the significance of each knob. By analyzing a given set of tuning observations, where each observation consists of a <configuration, performance metric> pair, the importance of each knob is determined through the calculation of its SHAP value. As highlighted in~\cite{Zhang22@tuning_evaluation}, SHAP currently stands out as the most effective learning method for assessing the importance of knobs. To gather training data for SHAP, we collect approximately 6000 observations for the SYSBENCH workload using the Latin Hypercube Sampling (LHS) method~\cite{McKay92@LHS}, which can sample configurations across the entire configuration space. Subsequently, we execute the workload under these configurations to acquire their corresponding performance metrics.

In addition, we invite an industry database expert to conduct the knob pruning task as a human annotation baseline. Specifically, we allow the expert to identify crucial database knobs and their important value ranges based on his own expertise and experience. The expert is also permitted to consult with other experienced database administrators, the broader database community, and official MySQL documentation to complete the task.

After narrowing the search space, we then utilize a traditional knob recommendation method, SMAC~\cite{Hutter11@SMAC}, to optimize these selected knobs for a maximum of 120 iterations. We record the configuration and corresponding database performance in each iteration to calculate the TES and ODP metrics.

\subsection{LLMs for Knob Pruning}
For LLMs, we utilize the prompt illustrated in Figure~\ref{fig:p_selection} to perform the knob pruning from the candidate knobs. It is worth noting that, during the inference of the LLM, we set the temperature parameter to 0 to guarantee deterministic outcomes.

\subsection{Experimental Results}
\begin{figure}
    \centering
    \includegraphics[width=0.45\textwidth]{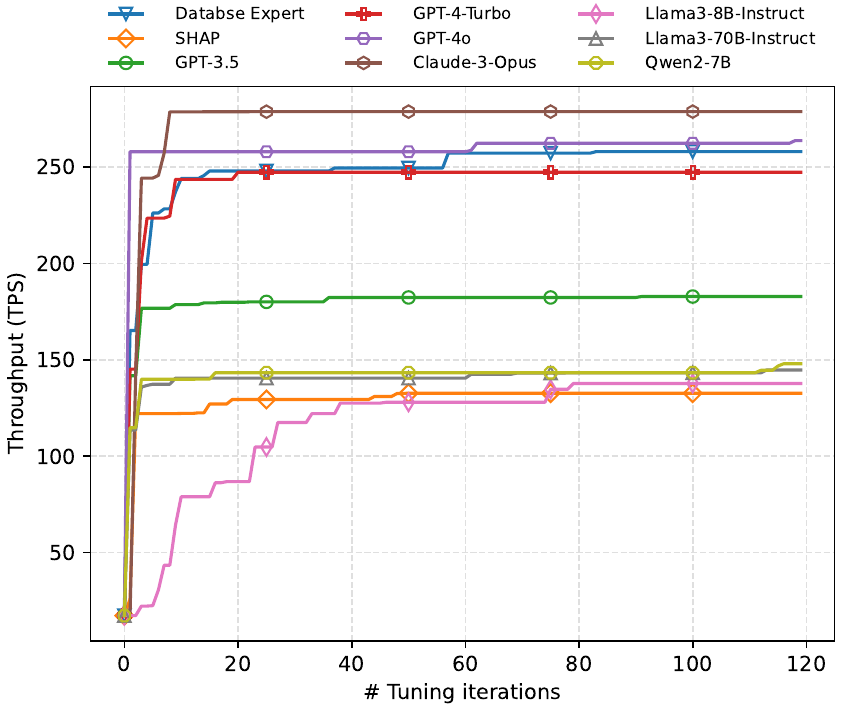}
    \caption{Best database performance over iterations. The horizontal axis represents the number of tuning iterations and the vertical axis represents the best TPS achieved (upper-left better). Different knob pruning methods result in different convergence speeds and optimal performance.} 
    \label{fig:res_selection}
\end{figure}

The experimental results are illustrated in Figure~\ref{fig:res_selection}. Our observations are as follows: (1) In the knob pruning task, certain LLMs (Claude-3-Opus, GPT-4o, and GPT-4-Turbo) demonstrate comparable or even superior performance to that of the database expert. Upon analysis, we discover a significant similarity between the knobs selected by the LLMs and those chosen by the expert. For instance, Table~\ref{tab:selected_knobs} showcases the top 10 most critical knobs identified by the database expert, GPT-4o, and SHAP. Notably, the knobs selected by GPT-4o closely align with those chosen by the database expert. This resemblance may be attributed to the extensive training data utilized for GPT-4o (and other LLMs), which incorporates MySQL community discussions, relevant articles, blogs, and official documentation~\cite{Zhou24@DBGPT}. Consequently, the LLM can emulate the behavior of the database expert, leading to similar knob pruning outcomes. (2) Furthermore, we note that nearly all evaluated LLMs outperform the previous learning-based method, SHAP, in terms of both convergence speed and optimal database performance, with GPT-4o exhibiting the most favorable results. After examining the selection outcomes produced by SHAP, we observe a distinct knob set compared to GPT-4o and the database expert.


\begin{table*}
\caption{Important knobs identified by the database expert, GPT-4o, and SHAP. In the comparison between the outcomes of GPT-4o and SHAP against those of the database expert, we emphasize the distinct selections by highlighting them in blue.}
\small
\centering
\setlength{\tabcolsep}{4pt} 
\renewcommand{\arraystretch}{1.05} 
\label{tab:selected_knobs}

\begin{tabular}{ccc}
    \toprule
    \textbf{Database Expert} 
&  \textbf{LLM (GPT-4o)} 
& \textbf{SHAP}  \\
    \midrule
    innodb\_buffer\_pool\_size 
&  innodb\_buffer\_pool\_size 
& innodb\_buffer\_pool\_size 
 \\
    tmp\_table\_size 
&  tmp\_table\_size& tmp\_table\_size 
 \\
     max\_heap\_table\_size 
& max\_heap\_table\_size  
& max\_heap\_table\_size \\
    innodb\_log\_file\_size 
& innodb\_log\_file\_size& \textcolor{blue}{\textbf{innodb\_compression\_failure\_threshold\_pct}} \\
    innodb\_flush\_log\_at\_trx\_commit 
& innodb\_flush\_log\_at\_trx\_commit& \textcolor{blue}{\textbf{query\_prealloc\_size}}\\
    query\_cache\_size 
& query\_cache\_size 
& \textcolor{blue}{\textbf{innodb\_thread\_concurrency}} \\
    table\_open\_cache 
& table\_open\_cache 
& \textcolor{blue}{\textbf{table\_open\_cache\_instances}}\\
    sort\_buffer\_size& \textcolor{blue}{\textbf{innodb\_io\_capacity}} 
& sort\_buffer\_size   \\
    max\_connections& \textcolor{blue}{\textbf{join\_buffer\_size}}& \textcolor{blue}{\textbf{innodb\_max\_dirty\_pages\_pct\_lwm}}\\
    key\_buffer\_size& \textcolor{blue}{\textbf{thread\_cache\_size}}& \textcolor{blue}{\textbf{innodb\_purge\_threads}} \\
  \bottomrule
\end{tabular}

\renewcommand{\arraystretch}{1.0}
\end{table*}


In delving into why GPT-4o surpasses the performance of the database expert in this task, a closer examination of their differences reveals a key distinction: as shown in Table~\ref{tab:selected_knobs}, GPT-4o opts for knob ``join\_buffer\_size'' while the database expert selects knob ``max\_connections''. Increasing ``join\_buffer\_size'' can enhance the efficiency of the join operator, consequently boosting overall database performance. On the other hand, the impact of increasing ``max\_connections'' on database performance is not always beneficial. If ``max\_connections'' exceeds the actual number of connections required by the workload, increasing the value of this knob will have no discernible effect on the database performance.

\subsection{Main Findings}
Our main findings of this section are summarized as follows:
\begin{itemize}[leftmargin=1em]
\setlength\itemsep{0em}
\item In the knob pruning task, certain LLMs (such as Claude-3-Opus and GPT-4o) demonstrate superior performance, surpassing even that of the DBAs. Furthermore, nearly all LLMs exhibit superior performance compared to the learning-based baseline, SHAP. 
\item The knobs chosen by some LLMs closely resemble those selected by the database expert, indicating the potential for LLMs to replace DBAs in automating the process of pruning knobs.
\item We observe closed-source LLMs are much better than open-source LLMs in this task, indicating that the source of the LLM can have a substantial impact on this task.
\item Simply prompting LLMs without fine-tuning any parameters can achieve performance levels comparable to that of humans, showcasing the remarkable flexibility and adaptability of LLMs in effectively addressing this task.
\item Therefore, one promising future direction is to fine-tune an LLM tailored for knob pruning, potentially enabling it to surpass experts by a considerable margin.
\end{itemize}

\section{Model Initialization}
\label{sec:transfer}
\subsection{Baselines}
To further enhance tuning efficiency, a series of model initialization methods~\cite{Aken17@OtterTune, Li19@qtune, Zhang21@ResTune} have been introduced to use the past tuning records to initialize learnable models in the knob recommendation methods.

Current model initialization methods can be broadly categorized into three main groups: workload mapping, model ensemble, and pre-training. Workload mapping, as proposed by OtterTune~\cite{Aken17@OtterTune}, involves matching the target workload with the most similar historical workloads and leveraging their tuning observations to initialize the surrogate model. This approach can be integrated into a wide range of BO-based knob recommendation methods. The model ensemble technique, as described in ResTune~\cite{Zhang21@ResTune}, entails collecting a set of well-established tuning models on historical workloads and then combining these models to guide the optimization of current tuning model for new workloads. Lastly, the pre-training technique is commonly utilized in RL-based knob recommendation methods, as seen in works such as QTune~\cite{Li19@qtune} and CDBTune~\cite{Zhang21@CDBTune}. This process involves initially pre-training parameters of the actor and the critic within the RL algorithm using a set of historical tuning records. Subsequently, when facing a new workload, the pre-trained models will be further fine-tuned. By avoiding the necessity to train models from randomly initialization, this technique could expedite the tuning process.


We have chosen representative methods from each category as baselines. For workload mapping, we have selected the OtterTune method~\cite{Aken17@OtterTune} integrated with the BO-based knob recommendation method VBO~\cite{Duan09@iTuned}. In the model ensemble category, we are using the ResTune method~\cite{Zhang21@ResTune} combined with the meta-learning knob recommendation method RGPE~\cite{Feurer2018@RGPE}. Lastly, for pre-training, we adopt QTune~\cite{Li19@qtune} as the baseline, which aims to accelerate the RL-based knob recommendation method DS-DDPG. In this section, all knob recommendation methods undergo 400 iterations with or without model initialization techniques.

\subsection{LLMs for Model Initialization}\label{sec:llm_for_kt}
We first use the prompt illustrated in Figure~\ref{fig:p_transfer} to sample 10 potentially effective configurations from LLMs. Subsequently, in the initial stages of VBO, we replace randomly sampled configurations with these 10 LLM-generated configurations to expedite its tuning process. To quantify the acceleration potential facilitated by LLMs, we consider the original VBO method for comparison. To generate a set of configurations from LLMs, we utilize nucleus sampling~\cite{holtzman2020@nucleus-sampling} with a temperature of 1.0 and a top-p value of 0.98. During our experimentation, we observe that certain sampled configurations are duplicated. As a result, we iteratively perform samplings until we acquire 10 distinct configurations.


\subsection{Experimental Results}

\begin{table}
\caption{Evaluation results for different model initialization methods. We report performance enhancement (\emph{i.e.}, PE) and speedup against the base model.}

\label{tab:transfer}
\begin{adjustbox}{width=0.48\textwidth}
\setlength{\tabcolsep}{4pt} 
\renewcommand{\arraystretch}{1.05} 

\begin{tabular}{cccccc}
    \toprule

     \multicolumn{1}{c}{\multirow{1}{*}{Type}} & \multicolumn{1}{c}{\multirow{1}{*}{Model}}     & \multirow{1}{*}{ODP}                        & \multirow{1}{*}{PE}  &\multirow{1}{*}{TES}   & \multirow{1}{*}{Speedup}
      \\
    \midrule

     \multirow{4}{*}{\makecell[c]{Traditional \\ Method}} & VBO & 154.73 &0\% & 316 & 0\% \\
     & VBO + Mapping & 154.37 & -0.23\% & 279 & 11.71\%  \\
     & RGPE + Model Ensemble & 158.02 & \textbf{0.42\%} & \textbf{215} & \textbf{42.67\%} \\
     & DS-DDPG + Pre-training & \textbf{162.15} & 33.10\% & 313 & -216.16\% \\
     \midrule
     \midrule

     \multirow{4}{*}{\makecell[c]{Closed Source\\ LLM}} & VBO + GPT-3.5& 127.68 & -17.48\% & 176 & 44.30\% \\
     & VBO + GPT-4-Turbo & \textbf{152.01} & \textbf{-1.93\%} & 84 & 73.41\%  \\
     & VBO + GPT-4o & 126.65 & -18.29\% & 90 & 71.51\%  \\
     & VBO + Claude-3-Opus & 126.09 & -18.65\% & \textbf{2} & \textbf{99.37\%} \\ 
     \midrule
     \midrule

     \multirow{3}{*}{\makecell[c]{Open Source\\ LLM}} & VBO + Llama3-8B-Instruct & 
153.16 &-1.19\% & \textbf{90} &  \textbf{71.51\%} \\
     & VBO + Llama3-70B-Instruct & \textbf{154.68} &\textbf{-0.03\%} & \textbf{90} & \textbf{71.51\%} \\
     & VBO + Qwen2-7B & 153.58 &-0.74\% & 100 & 68.35\%\\

  \bottomrule
\end{tabular}

\end{adjustbox}
\renewcommand{\arraystretch}{1.0}
\end{table}

We present the experimental results in Table~\ref{tab:transfer}. To provide a more intuitive understanding of the effectiveness of different initialization methods, we adopt the approach outlined in~\cite{Zhang22@tuning_evaluation} to introduce two additional metrics: performance enhancement (PE) and speedup. Specifically, we represent the TES and ODP values for the base knob recommendation method without initialization as $TES_{orig}$ and $ODP_{orig}$, and for the method with initialization as $TES_{init}$ and $ODP_{init}$. In the OLTP benchmark, a higher TPS value represents better performance. Therefore, the performance enhancement is calculated as follows:
\begin{equation}
    PE = \frac{ODP_{init} - ODP_{orig}}{ODP_{orig}},
\end{equation}
and the speedup is defined as follows:
\begin{equation}
    Speedup = \frac{TES_{orig} - TES_{init}}{TES_{orig}}.
\end{equation}

The PE metric assesses whether the initialization technique can aid the base knob recommendation method in identifying superior configurations. Subsequently, the speedup metric measures the degree to which the initialization technique expedites the tuning process. Higher values for both PE and speedup indicate improved performance of an initialization method.

The experimental findings are detailed in Table~\ref{tab:transfer}. Initially, we observe that the workload mapping technique yields a modest speedup of 11.71\%. This outcome could be attributed to the limited utilization of historical tuning records. Moving on to the model ensemble technique, despite delivering a substantial 42.67\% speedup, it still necessitates 215 iterations to reach peak performance. The outcomes of the pre-training technique are surprising. While it does lead to a significant performance improvement of 33.10\%, it also causes a drastic decrease in speedup by -216.16\%, a result that is unacceptable. In summary, these conventional initialization techniques do not clearly expedite the tuning process, which still necessitates 200-300 iterations to achieve the optimal performance.


For using LLMs in the model initialization task, we first observe that some LLMs (GPT-3.5, GPT-4o, and Claude-3-Opus) exhibit relatively poor performance enhancement and largely under-perform the base model (\emph{i.e.}, VBO). After analyzing the sampling configurations of these LLMs, we find that despite providing 20 knobs in the prompt, these models predominantly adjust only a handful of knobs, leaving the rest at default settings. Consequently, the 10 LLM-generated configurations exhibit significant similarities. Utilizing these configurations to initialize VBO might limit its capacity to explore uncharted areas and thus result in poor performance enhancement. To address this issue, one approach is to increase the temperature value to inject more randomness during the sampling phase. Among the other LLMs, including GPT-4-Turbo, Llama3-8B-Instruct, Llama3-70B-Instruct, and Qwen2-7B, a slight performance decrease (less than 2\%) is observed, falling within an acceptable range. Moreover, initializing VBO with LLM-generated configurations leads to a noticeable acceleration in convergence speed. Notably, LLMs have shown an average speedup of 71.42\%, with some achieving an impressive 99.37\% boost. In essence, utilizing LLMs for initializing VBO requires a careful balance between ODP and TES. Given the minor performance impact and significant tuning acceleration, this trade-off is deemed acceptable.



\subsection{Main Findings}
Our main findings of this section are summarized as follows:
\begin{itemize}[leftmargin=1em]
\setlength\itemsep{0em}
\item Existing model initialization methods do not demonstrate an obvious speedup in the tuning process, often necessitating hundreds of iterations to identify a suitable configuration. On the other hand, LLMs have shown their capability to greatly expedite the convergence of BO-driven knob recommendation approaches through the generation of initial configurations. 
\item In contrast to the findings in knob pruning, we note that open-source LLMs outperform closed-source LLMs in this particular task. This disparity arises from the tendency of closed-source LLMs to frequently generate similar configurations, thereby constraining the exploration and exploitation capacities of the subsequent BO-based knob recommendation method. 
\item Hence, a promising direction for future exploration lies in leveraging LLMs to generate a variety of valuable and distinct configurations to initialize the base knob recommendation methods.

\end{itemize}

\section{Knob Recommendation}
\label{sec:rec}
\subsection{Baselines}
The knob recommendation phase stands out as the core component of the entire tuning system, exerting a direct influence on the final tuning performance. Within this section, we evaluate LLMs' knob recommendation capability against three widely-used traditional knob tuners: Vanilla Bayesian Optimization (VBO)~\cite{Duan09@iTuned}, Sequential Model-based Algorithm Configuration (SMAC)~\cite{Hutter11@SMAC}, and Deep Deterministic Policy Gradient (DDPG)~\cite{Lillicrap15@DDPG}. VBO is a BO-based konb recommendation method utilizing a vanilla Gaussian Process (GP) as its surrogate model. The vanilla GP aims to model the relationship between the configuration and the database performance~\cite{Duan09@iTuned, Aken17@OtterTune}. On the other hand, SMAC is another BO-based approach that utilizes a random forest algorithm as its surrogate model to guide the tuning process~\cite{Breiman01@RF}. 
DDPG is an RL-based method that has been widely integrated into existing knob tuning frameworks. It distinguishes itself from traditional RL algorithms like deep Q-learning~\cite{Maglogiannis18@QLearning} by its ability to operate in both discrete and continuous action spaces.
DDPG involves the training of two neural networks: the actor network, responsible for selecting actions (\emph{i.e.}, configurations) based on database states, and the critic network, which evaluates the chosen action's reward (\emph{e.g.}, latency or transactions per second). This reward signal updates the actor network, enabling it to make better decisions in subsequent iterations. Following the settings used in Section~\ref{sec:llm_for_kt}, all these base knob tuners undergo 400 iterations.






\subsection{LLMs for Knob Recommendation}
As outlined in Section~\ref{sec:problem definition}, we frame the knob recommendation task for LLMs as an ``iteratively refining'' process. Specifically, starting from the default configuration, we iteratively refine the current configuration using feedback from the database. In our evaluation, we also introduce a new metric called ``IR TPS'' to measure the database performance after the Initial Refinement step. Considering the high cost associated with using closed-source models, we restrict the LLMs to undergo a maximum of 30 rounds of refinement. Furthermore, we set the temperature to 0 to generate deterministic LLM outputs.

\subsection{Experimental Results}
\label{sec:recommend result}

\begin{table}
\caption{Comparing database performance and tuning efficiency among various knob recommendation methods. ``Default'' denotes the use of the default configuration.}
\small
\label{tab:knob recommendation}

\begin{tabular}{ccccc}
    \toprule

     Type & Method & IR TPS & ODP & TES     \\
    \midrule

    \multirow{4}{*}{\makecell[c]{Traditional\\ Method}} & Default & - & 17.45  & -\\
    & DDPG  & - & 120.71 &  \textbf{99} \\
    & SMAC & - & \textbf{157.25}  & 375 \\
    & VBO & - & 155.10  & 316 \\
    \midrule
    \midrule

    \multirow{4}{*}{\makecell[c]{Closed Source\\ LLM}} & GPT-3.5 &  16.38 & 116.62  & 24 \\
    & GPT-4-Turbo & \textbf{145.06} & \textbf{155.30}  & \textbf{9} \\
    & GPT-4o & 117.96 & 126.05  & 13 \\
    & Claude-3-Opus & 26.70 & 148.51 & 23 \\ 
    \midrule
    \midrule

    \multirow{3}{*}{\makecell[c]{Open Source\\ LLM}} & Llama3-8B-Instruct & 20.90 & 125.88  & 25 \\
    & LlaMa3-70B-Instruct & 28.84 & 145.12  & \textbf{3} \\
    & Qwen2-7B & \textbf{143.58} & \textbf{154.94} & 14 \\

  \bottomrule
\end{tabular}

\end{table}

We present experimental results in Table~\ref{tab:knob recommendation}. In the realm of traditional methods, two BO-based strategies, SMAC and VBO, surpass the singular RL-based method, DDPG, in terms of ODP, leading to enhanced database performance. Nevertheless, in terms of TES, both BO-based methods require over 300 iterations, a considerable increase compared to DDPG, which achieves convergence in just 99 iterations.

For LLMs in the knob recommendation task, we observe remarkable results. Specifically, it's impressive that some LLMs, including GPT-4-Trubo, GPT-4o, and Qwen2-7B, show a great performance enhancement over the default configuration within only one-step refinement (see ``IR TPS''). This finding reveals that the LLM has the potential to be an end-to-end knob recommendation method without extensive iterations required in existing methods. Furthermore, leveraging feedback from the database, LLMs can iteratively suggest enhanced configurations, attaining comparable database performance to traditional methods but with significantly fewer iterations. Notably, in the instances of GPT-4-Turbo and Qwen2-7B, their refined configurations closely rival the top-performing traditional method, SMAC (155 versus 157), yet GPT-4-Turbo and Qwen2-7B achieve optimal performance in just 9 and 14 iterations, respectively, compared to SMAC's 375 iterations. These findings indicate that LLMs possess the capability to comprehensively understand and utilize database feedback to refine configurations for enhanced overall database performance, showcasing remarkable exploration and exploitation capabilities.



\begin{figure}
    \centering
    \includegraphics[width=0.48\textwidth, ]{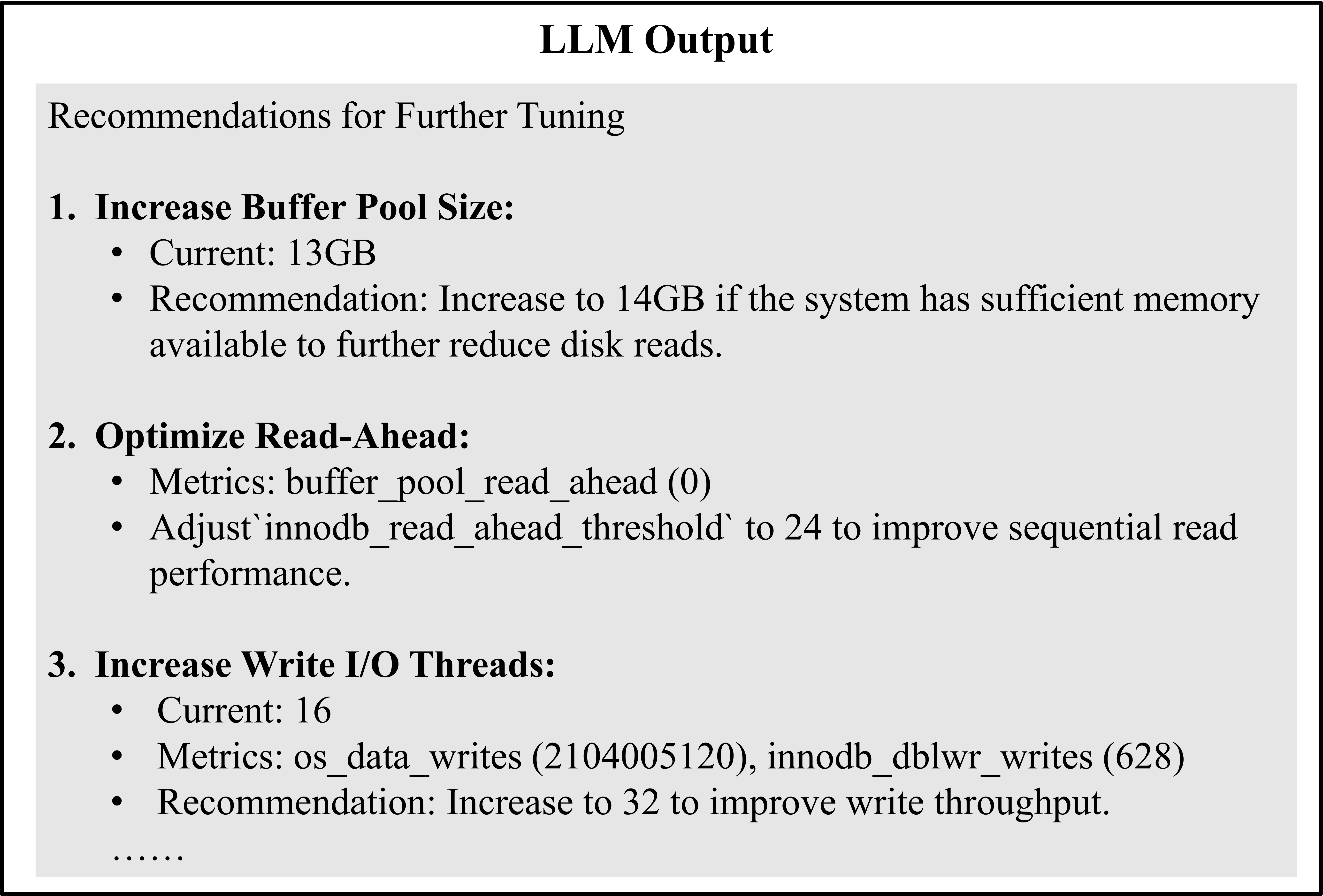} 
    \caption{Illustration of tuning suggestions offered by LLMs.} 
    \label{fig:LLM output}
\end{figure}

Furthermore, LLMs demonstrate better interpretability in contrast to traditional black-box approaches. Illustrated in Figure~\ref{fig:LLM output}, when assigned the role of configuration recommendation, the LLM consistently provides detailed rationales and considerations in the ``chain-of-thought''~\cite{wei2022@cot} manner for each adjustment made to the knobs. This attribute not only bolsters practical applicability but also enhances the reliability of the results. On the other hand, by presenting the reasoning behind its recommendations, LLMs can help DBAs in making informed decisions, fostering a collaborative and streamlined process of knob recommendation.


\begin{figure}
    \centering
    \includegraphics[width=0.45\textwidth]{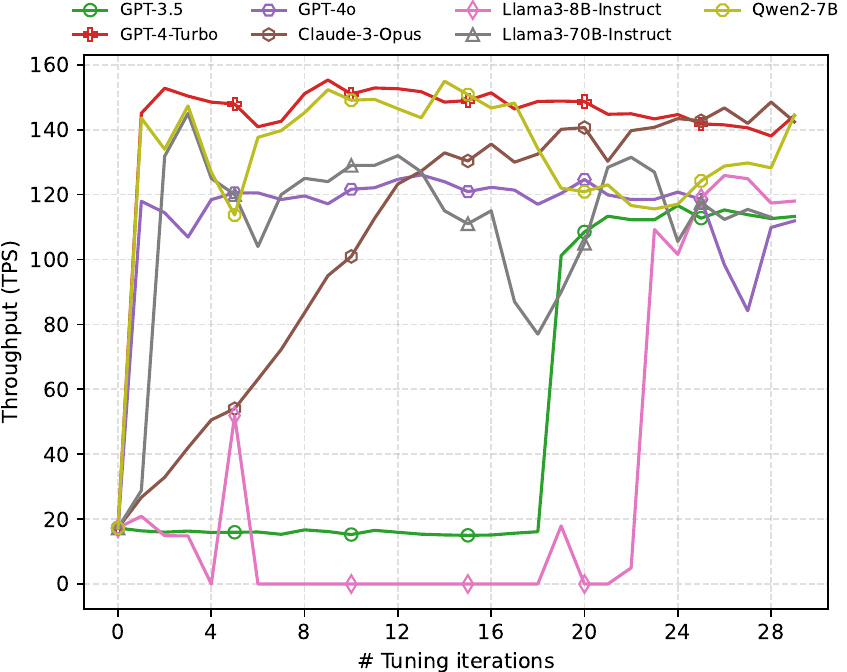}
    \caption{The impact of iteration count on database performance. The x-axis displays the number of iterations, while the y-axis represents the database performance achieved through refined configurations after each iteration. Higher values on the y-axis represent better performance levels.} 
    \label{fig:iterative LLM}
\end{figure}

Finally, by comparing ``IR TPS'' and ODP metric in Table~\ref{tab:knob recommendation}, we can observe that the ``iterative refinement'' strategy can significantly improve the quality of the recommended configurations for some LLMs. Take Claude-3-Opus as an example: initially, the first refinement yields only 26 TPS. However, through the iterative refinement process, a significantly superior configuration yielding 148 TPS is identified. To examine the correlation between the number of iterations and database performance, we present the findings in Figure~\ref{fig:iterative LLM}. Our analysis reveals that nearly all LLMs effectively leverage the database feedback to suggest improved configurations in the iteration process.


\subsection{Main Findings} 
Our main findings of this section are summarized as follows:
\begin{itemize}[leftmargin=1em]
\setlength\itemsep{0em}
    \item In terms of tuning effectiveness, LLMs demonstrate a remarkable ability to achieve database performance on par with traditional methods. Moreover, LLMs can pinpoint promising configurations in significantly fewer iterations, showcasing impressive tuning efficiency.
    \item Contrasted with traditional black-box methods, the knobs suggested by LLMs offer enhanced interpretability. For DBAs, these LLM-recommended knobs are more ``traceable'', facilitating further adjustments based on their expert knowledge.
    \item We do not observe a notable performance difference between open-source and closed-source LLMs in the knob recommendation task. Both variants demonstrate similar effectiveness.
    \item The development of an end-to-end LLM-based knob recommendation approach could represent a promising research avenue. Such a method has the potential to eliminate the need for extensive iterations present in previous methods.



\end{itemize}

\section{Generalizability of LLMs}
\label{sec: generalization}
In this section, we delve into the generalizability of LLMs by expanding our evaluation framework to encompass new benchmarks, database engines, and hardware environments. As detailed in Section~\ref{sec:setup}, our primary evaluations are conducted using the SYSBENCH benchmark, the MySQL engine, and a server with 8 CPUs and 16 GB of RAM. Therefore, we can modify a single variable—keeping the others constant—to assess the LLMs' adaptability and performance across diverse scenarios. For example, transitioning from the SYSBENCH to the JOB benchmark~\cite{Leis15@JOB}, switching from MySQL to PostgreSQL, or changing from one server configuration to another allows us to create new evaluation contexts.

Traditional methods, when confronted with new evaluation environments, typically require substantial code modifications, exhaustive data collections, and comprehensive re-training. These processes are not only time-consuming but also demand significant labor and resources. In contrast, LLMs offer a distinct advantage due to their prompt-driven nature; modifying prompts can enable smooth transitions across diverse evaluation setups.

As previously outlined, knob recommendation is central to the tuning system and directly impacts database performance. This section, therefore, focuses on the proficiency of LLMs in the task of knob recommendation. Given the flexibility of LLMs, tasks such as knob pruning and model initialization can also be efficiently completed by adjusting the prompts provided to the LLMs for new evaluation scenarios.

\subsection{Setup}
To establish new evaluation setups, we consider modifications across three key dimensions: benchmark, database engine, and machine instance.

\vpara{Benchmark.} 
To broaden our evaluation of LLMs' knob recommendation abilities to analytical workloads, we introduce an OLAP benchmark named JOB~\cite{Leis15@JOB}. JOB comprises 113 benchmarked queries featuring intricate joins and a database instance storing 9GB of data. We consolidate all 113 queries into a single OLAP workload.

\vpara{Database Engine.}
To assess the versatility of LLMs across different database engines, we have chosen two famous open-source database systems: PostgreSQL and TiDB~\cite{Huang20@TiDB}. Initially released in 1996, PostgreSQL stands out as a widely adopted relational database recognized for its advanced functionalities, reliability, and adeptness in handling complex queries. We are confident that a wealth of manuals and online forums dedicated to PostgreSQL exist on the internet, and LLMs have extensively pre-trained on this content to assimilate domain knowledge related to PostgreSQL. For our experiments, we employ PostgreSQL version 10.0, carefully choosing 20 crucial knobs for optimization, while maintaining the default values for others as specified in the configuration file.

On the other hand, TiDB is a new database engine (released in 2017) that focuses on the distributed database system, supporting Hybrid Transactional and Analytical Processing (HTAP) workloads~\cite{Huang20@TiDB}. Given that TiDB is a relatively new player in the database engine landscape, with limited discussion available online, it offers an opportunity to assess the adaptability of LLMs to newer database engines. In practice, we use TiDB version 8.2 and instantiate the TiDB cluster with a complete topology, simulating the production deployment. This setup includes 3 TiKV instances (row-based storage engine), 1 TiFlash instance (column-based storage engine), 1 TiDB instance (distributed database server), 1 PD instance (placement driver), and 1 Monitor. Following the official documentation, we pinpoint 7 important knobs for tuning, leaving the remainder at their automatic, hardware-configured default settings. This contrasts with PostgreSQL, where TiDB's default knob values adjust dynamically to match the hardware specifications.

\vpara{Hardware Environment.}
To explore the generalizability at the hardware level, we have deployed the MySQL database on a distinct machine featuring 40 CPUs and 256 GB of RAM. For clarity, we will refer to the prior machine with 8 CPUs and 16GB of RAM as machine A, and designate this new machine with 40 CPUs and 256GB of RAM as machine B.

\subsection{Experimental Results}
\subsubsection{Evaluation on Varying Benchmark.}
We have substituted SYSBENCH with an OLAP benchmark JOB and tasked LLMs with recommending a configuration to minimize the latency of the specified workload. The results of our experiments are detailed in Table~\ref{tab:general_workload}. In this table, ``IR Latency'' denotes the latency of the workload (in seconds) under the configuration produced through the initial refinement step. ``ODP$_{AP}$'' represents the minimum latency achieved during the whole refinement process. 

After analyzing the results, the following key findings emerge: (1) Almost all LLMs are able to identify appropriate configurations after the initial refinement, highlighting the potential of LLMs as an end-to-end knob recommendation solution. (2) Moreover, through iterative refinement, GPT-4-Turbo and Llama3-70B-Instruct outperform the leading traditional method, SMAC, by identifying comparable or superior configurations on the JOB benchmark. (3) Lastly, LLMs demonstrate impressive tuning efficiency, typically requiring only a few iterations to achieve promising configurations. We should emphasize that, in the context of an OLAP workload, enhanced tuning efficiency is more critical compared to an OLTP workload. For OLTP, we evaluate a configuration's performance through a stress test over a fixed period. In contrast, OLAP tuning involves evaluating a configuration's performance by executing all SQL statements in a workload. This can lead to significantly high time costs, especially when encountering slow SQL queries, resulting in an extremely long evaluation time for a tuned configuration. Therefore, OLAP workloads necessitate more efficient tuning strategies.

These findings indicate that even after transitioning the benchmark from OLTP to OLAP, LLMs can consistently recommend top-tier configurations by effectively understanding and incorporating the characteristics of the OLAP benchmark.

\subsubsection{Evaluation on Varying Database Engine.}
After substituting MySQL with PostgreSQL and TiDB, we present the experimental results in Table~\ref{tab:general_database}. Specifically, for PostgreSQL, we observe similar results on MySQL: Compared to the default configuration, both traditional methods and LLMs can find much better configurations. In addition, LLMs only require 1\%-10\% steps of traditional methods while finding comparable configurations. 

For the newer database engine TiDB, its capability to automatically set default knob values based on hardware specifications significantly contributes to its superior performance over PostgreSQL in the ``Default'' setting, with TiDB achieving 164.00 TPS compared to PostgreSQL's 52.00 TPS. This automatic optimization presents a notable challenge for knob recommendation methodologies, as the baseline performance is already optimized. As illustrated in Table~\ref{tab:general_database}, configurations recommended by the DDPG algorithm even fall short of TiDB's default settings. In addition, the initial refinement of GPT-3.5, GPT-4o, and Qwen2-7B also lead to a decline in database performance. Nonetheless, through iterative refinement and leveraging database feedback, all LLMs eventually surpass the default configurations. Remarkably, Claude-3-Opus outperforms the traditional method SMAC in terms of ODP with a score of 265.71 versus 263.53 and demonstrates greater efficiency in TES, requiring only 21 iterations compared to SMAC's 342 iterations.

These findings underscore the robustness of LLMs across diverse database engines. By presenting knob descriptions and inner metric details in the prompt, LLMs leverage their internal knowledge and linguistic comprehension ability acquired during pre-training to extrapolate to unfamiliar database engines.

\subsubsection{Evaluation on Varying Hardware Environment.}
After migrating our MySQL database to a more powerful server (referred to as machine B), we detail the experimental outcomes in Table~\ref{tab:general_hardware}. The enhanced CPU capabilities and increased RAM on machine B necessitate a broader range for certain knobs, thereby complicating the task of pinpointing the optimal configuration. The experimental data reveal that LLMs significantly outperform traditional knob recommendation methods on this upgraded hardware. Notably, the most effective traditional method, SMAC, achieves a database performance of 1343.68 TPS, whereas GPT-4-Turbo, GPT-4o, Claude-3-Opus, and Qwen2-7B, exceed this benchmark with considerably fewer iterations, achieving 1546.26, 1462.42, 2424.77, 1450.23 TPS, respectively.

These findings underscore the adaptability of LLMs to hardware modifications, attributed primarily to their comprehensive understanding of the relationship between knob settings and hardware specifications, which effectively circumvents these challenges.





\begin{table}
\caption{Experimental results of ``JOB + MySQL + machine A (8 CPUs and 16GB RAM)''. Notably, for ``IR Latency'' and ``ODP$_{AP}$'', which indicate the latency of the JOB benchmark, lower values signify improved performance."
}
\small
\label{tab:general_workload}

\begin{tabular}{ccccc}
    \toprule
     Type & Method & IR Latency & ODP$_{AP}$ & TES     \\
    \midrule

    \multirow{4}{*}{\makecell[c]{Traditional\\ Method}} & Default & - & 2594.27  & -\\
    & DDPG  & - & 853.11 &  179 \\
    & SMAC & - & \textbf{675.28}  & 146 \\
    & VBO & - & 716.90  & \textbf{116} \\
    \midrule
    \midrule

    \multirow{4}{*}{\makecell[c]{Closed Source\\ LLM}} & GPT-3.5 &  812.92 & 757.98  & 19 \\
    & GPT-4-Turbo & \textbf{712.38} & \textbf{643.77}  & 7 \\
    & GPT-4o & 852.29 & 833.36  & \textbf{3} \\
    & Claude-3-Opus & 829.15 & 794.95 & 12 \\ 
    \midrule
    \midrule

    \multirow{3}{*}{\makecell[c]{Open Source\\ LLM}} & Llama3-8B-Instruct & 932.64 & 799.72  & 9 \\
    & LlaMa3-70B-Instruct & 829.12 & \textbf{673.23}  & \textbf{7} \\
    & Qwen2-7B & \textbf{799.63} & 713.19 & 12 \\

  \bottomrule
\end{tabular}

\end{table}
\begin{table}
\caption{Experimental results of ``SYSBENCH + PostgreSQL / TiDB + machine A (8 CPUs and 16GB RAM)''.}
\begin{adjustbox}{width=0.48\textwidth}
\label{tab:general_database}

\begin{tabular}{cccccccc}
    \toprule

     \multicolumn{1}{c}{\multirow{2}{*}{Type}} & \multicolumn{1}{c}{\multirow{2}{*}{Method}} & \multicolumn{3}{c}{PostgreSQL} & \multicolumn{3}{c}{TiDB} \\ 
     \cmidrule(lr){3-5} \cmidrule(lr){6-8}
      &  &  IR TPS & ODP & TES & IR TPS & ODP & TES    \\
    \midrule

    \multirow{4}{*}{\makecell[c]{Traditional\\ Method}} & Default & - & 52.00  & - & - & 164.00 & - \\
    & DDPG  & - & 133.75 &  193 & - & 109.59 & \textbf{311} \\
    & SMAC & - & \textbf{147.03}  & 189 & - & \textbf{263.53} & 342 \\
    & VBO & - & 140.52  & \textbf{116} & - & 253.01 & 355\\
    \midrule
    \midrule

    \multirow{4}{*}{\makecell[c]{Closed Source\\ LLM}} & GPT-3.5 &  \textbf{136.32} & 136.32  & \textbf{1} & 156.21 & 180.89 & 10\\
    & GPT-4-Turbo & 126.88 & \textbf{143.77}  & 12 & 218.43 & 255.73 & 5 \\
    & GPT-4o & 68.95 & 142.36  & 19  & 110.69 & 218.20 & \textbf{4}\\
    & Claude-3-Opus & 101.73 & 127.95 & 5  & \textbf{263.43} & \textbf{265.71} & 21 \\ 
    \midrule
    \midrule

    \multirow{3}{*}{\makecell[c]{Open Source\\ LLM}} & Llama3-8B-Instruct & \textbf{124.82} & 144.41  & 11 & 165.05 & \textbf{221.13} & 8\\
    & LlaMa3-70B-Instruct & 120.10 & \textbf{152.68}  & \textbf{3} & \textbf{179.88} & 213.03 & \textbf{3}\\
    & Qwen2-7B & 100.94 & 120.35 & 13 & 153.75 & 198.62 & 17\\

  \bottomrule
\end{tabular}

\end{adjustbox}
\end{table}
\begin{table}
\caption{Experimental results of ``SYSBENCH + MySQL + machine B (40 CPUs and 256GB)''.}
\small
\label{tab:general_hardware}

\begin{tabular}{ccccc}
    \toprule
     Type & Method & IR TPS & ODP & TES     \\
    \midrule

    \multirow{4}{*}{\makecell[c]{Traditional\\ Method}} & Default & - & 500.42  & -\\
    & DDPG  & - & 1129.01 &  \textbf{210} \\
    & SMAC & - & \textbf{1343.68}  & 287 \\
    & VBO & - & 1271.61  & 371 \\
    \midrule
    \midrule

    \multirow{4}{*}{\makecell[c]{Closed Source\\ LLM}} & GPT-3.5 &  1280.16 & 1287.91  & 7 \\
    & GPT-4-Turbo & 1473.48 & 1546.26  & \textbf{2} \\
    & GPT-4o & 1360.93 & 1462.42  & 13 \\
    & Claude-3-Opus & \textbf{2050.79} & \textbf{2424.77} & 3 \\ 
    \midrule
    \midrule

    \multirow{3}{*}{\makecell[c]{Open Source\\ LLM}} & Llama3-8B-Instruct & 732.26 & 1099.31  & 19 \\
    & LlaMa3-70B-Instruct & 1007.37 & 1021.16  & \textbf{5} \\
    & Qwen2-7B & \textbf{1317.86} & \textbf{1450.23} & 9 \\

  \bottomrule
\end{tabular}

\end{table}

\subsection{Main Findings}
\begin{itemize}[leftmargin=1em]
\setlength\itemsep{0em}
    \item Compared to conventional approaches, LLMs consistently deliver stable and commendable results across various benchmarks, database engines, and hardware configurations. Furthermore, employing LLMs in these new evaluation contexts requires minimal code adjustments due to their flexibility through prompting techniques. This insight highlights the potential of LLMs in navigating more complex and dynamic tuning environments and tasks.
    
    \item Our observations indicate that, in most new evaluation settings, closed-source LLMs outperform their open-source counterparts, suggesting superior generalization capabilities in closed-source models.
    
    \item LLMs demonstrate proficiency with well-established database engines like MySQL and PostgreSQL. In the case of newer database engines such as TiDB, LLMs can effectively adapt by incorporating specific knob and metric information within the prompts, facilitating a smooth transition to unfamiliar environments.

    \item Among the LLMs evaluated, GPT-4-Turbo and Claude-3-Opus consistently exhibit the most reliable performance. Given their proprietary nature, a promising direction for future work involves creating an open-source LLM that matches or exceeds the capabilities of these closed-source models.

\end{itemize}

\section{discussion}
\label{sec:discussion}
In this section, we summarize our main findings in this work and explore potential avenues for future research. 

\subsection{Main Findings}


\vpara{Finding 1:} LLMs have demonstrated impressive abilities in enhancing database management across three pivotal subtasks in knob tuning: knob pruning, model initialization, and knob recommendation. Particularly in the knob recommendation task, LLMs can identify comparable or superior configurations with only a few iterations, showcasing their tuning efficiency.


\vpara{Finding 2:} We have noted that LLMs consistently employ a ``chain-of-thought'' approach in generating responses across various subtasks. Consequently, leveraging LLM-based solutions can greatly improve interpretability in addressing knob-tuning-related challenges

\vpara{Finding 3:} Diverse LLMs exhibit varying levels of performance across three knob tuning subtasks, with no single LLM consistently surpassing the others. This phenomenon can likely be attributed to the differences in the training corpora among various LLMs. Nevertheless, when considering closed-source LLMs, it is evident that GPT-4-Turbo delivers superior performance across various scenarios. On the other hand, for open-source LLMs, Llama3-70B-Instruct emerges as a commendable option.



\vpara{Finding 4:} LLMs demonstrate impressive generalizability across diverse evaluation scenarios, encompassing new benchmarks, database kernels, and hardware environments. This capability necessitates no further training, merely the adjustment of prompts.


\subsection{Research Opportunities}
\vpara{Opportunity 1:} For the task of knob pruning, we can fine-tune an LLM to better capture the relationship between the workload and its important knobs. In addition, we can also use the retrieval-augmented generation (RAG) technique to efficiently utilize more tuning experiences from the website.

\vpara{Opportunity 2:} For the task of model initialization, finding a way to sample diverse and useful configurations from LLMs would be a promising direction.

\vpara{Opportunity 3:} For knob recommendation, we can use LLMs to perform end-to-end knob recommendation by designing a more comprehensive prompting pipeline or fine-tuning an LLM using a high-quality training set containing numerous <workload, optimal configuration> data pairs.

\vpara{Opportunity 4:} As we mentioned above, no single LLM consistently outperforms the others in our evaluations. Therefore, a promising research direction is to design a method that dynamically selects appropriate LLM for the given evaluation setup.

\vpara{Opportunity 5:} Finally, creating a pre-trained LLM specifically designed for database management could significantly benefit the database community. To achieve this, we could gather a substantial corpus related to databases from the web and further pre-train an LLM, thereby infusing it with domain-specific knowledge.




\section{Conclusion}
\label{sec:conclusion}
Large Language Models (LLMs) have proven their efficacy and resilience across an extensive array of natural language processing tasks. In this paper, we conduct thorough experiments to explore LLMs' capabilities in the context of database knob tuning. We decompose the tuning system into three distinct subtasks: knob pruning, model initialization, and knob recommendation, and then use LLMs to complete each of them. To accomplish this, we transform each subtask into a sequence-to-sequence generation task and meticulously design their respective prompt templates. Compared to conventional state-of-the-art techniques, LLMs not only exhibit superior performance in these areas but also show remarkable interpretability by generating suggestions in a chain-of-thought manner. Furthermore, we conduct a variety of experiments to assess the adaptability of LLMs under different evaluation setups, such as changing workloads, database engines, and hardware configurations. We hope that this study will not only advance the field of knob tuning but also encourage further AI-driven tasks in databases, including query optimization, index recommendation, and more.

\bibliographystyle{ACM-Reference-Format}
\bibliography{sample}

\end{document}